\pdfoutput=1
\documentclass[10pt,twocolumn]{confpaper}

\graphicspath{ {./figures/} }

\date{}
\title{\ttlfnt{
\TITLE}}
\author{
\aufnt{\AUTHORS} \\
\affaddr{Princeton University}
}

\begin{document}
\thispagestyle{empty}

\maketitle


\begin{sloppypar}

\begin{abstract}
The proliferation of networked devices, systems, and applications that
we depend on every day makes managing networks more
important than ever. The increasing security, availability, and performance demands
of these applications suggest
that these increasingly difficult network management problems be solved in real time,
across a complex web of interacting protocols and systems.
Alas, just as the importance of network management has increased, the network
has grown
so complex that it is seemingly unmanageable. 
In this new era, network management
requires a fundamentally new approach. 
Instead of optimizations
based on closed-form analysis of individual
protocols, network operators need data-driven, machine-learning-based models of
end-to-end and application performance
based on high-level policy goals and a holistic view of the underlying  components. 
Instead of anomaly
detection algorithms that operate on offline analysis of network traces, operators
need classification and detection algorithms that can make real-time, closed-loop
decisions. 
Networks should learn to drive themselves.
This paper explores this concept, discussing how we might attain this ambitious
goal by more
closely coupling measurement with real-time control and by relying on learning for inference and prediction about a networked application or system, as
opposed to closed-form analysis of individual protocols.
\end{abstract}

\ifthenelse{\equal{\onlyAbstract}{no}}{

\section{Introduction}

Modern networked applications operate at a scale and scope we have never seen
before. Virtual and augmented reality require real-time responsiveness,
micro-services deployed using containers introduce rapid changes in traffic
workloads, and the Internet of Things (IoT) significantly increases the number
of connected devices while also raising new security and privacy concerns. The
widespread integration of these applications into our daily lives raises the
bar for network management, as users elevate their expectations for real-time
interaction, high availability, resilience to attack, ubiquitous access, and
scale. Network management has always been a worthwhile endeavor, but now it is
mission critical.

Yet, network management has remained a Sisyphean task. Network operators
develop and use scripts and tools to help them plan, troubleshoot, and secure
their networks, as user demands and network complexity continue to grow.
Networking researchers strive to improve the tuning, design, and measurement of
network protocols, yet we continue to fall behind the curve, as the protocols,
variable network conditions, and relationships between them and user quality
of experience become increasingly complex. 
Twenty years ago, we had some hope of (and success in) creating clean, closed-form
models of 
individual protocols, applications, and
systems~\cite{arlitt2005,padhye1998}; today, many of these are too
complicated for closed-form analysis. Prediction problems such as determining
how search query response time would vary in response to the placement of a
cache are much more suited to statistical inference and machine learning based on measurement data~\cite{tariq2008}.
 
Of course, we must change the network to make network management easier. We have
been saying this for years, as we continue to fall behind the curve. Part of
the problem, we believe, is the continued
focus on designing, understanding, and tweaking individual
protocols---we focus on better models for BGP, optimizations for TCP, QUIC,
DNS, or the protocols du jour. In fact, our troubles do not lie in the
protocols. The inability to model holistic network systems, as opposed to
individual protocols, has made it difficult for operators to understand what
is happening in the network. Software-Defined Networking
(SDN) helps by 
offering greater programmability and centralized control, yet controller
applications still rely on collecting their own data and installing low-level match-action
rules in switches and
SDN does not change the fact that real networked systems are too complex
to analyze
with closed-form models.

As networking researchers, we must change our approach to these problems.
An ambitious goal for network management is that of a {\em self-driving network}---one where (1)~network measurement
is task-driven and tightly integrated with the control of the network; and
(2)~network control relies on learning and large-scale data analytics of
the entire networked system, as opposed to closed-form models of individual
protocols. Recent initiatives have proffered this high-level goal~\cite{juniper-sdn,stanford-sdn}, drawing an analogy to self-driving cars, which can make decisions that manage uncertainty and mitigate
risk to achieve some task (\eg, transportation to some destination). This paper explores this goal in detail, developing the technical requirements for and properties of a self-driving network and outlining a broad, cross-disciplinary research agenda for the community that can move us closer to realizing this goal.


The networking research community has been developing the pieces of this puzzle
for many years,
from predictive models of application performance~\cite{tariq2008,li2010} to statistical anomaly and
intrusion detection algorithms based on analysis of network traffic~\cite{antonakakis2010,gu2008}. The state
of the art, however, merely lays the foundation for the much more ambitious agenda
of creating a truly self-driving network. Today, measurement remains decoupled from
network control, inevitably placing the network operator in the middle of the
control loop and introducing uncertainty and the possibility for error. Taking
the technologies that we have and making them both real-time and distributed
introduces entirely new classes of challenges, in networking and more broadly across
computer science:

\paragraph{Deriving measurement, inference, and control from high-level policy}:
A self-driving network should take as input a high-level goal related to
(say) performance or security and {\em jointly derive} (1)~the measurements that
the network should collect, (2)~the inferences that should be performed, and (3)~the decisions
that the network should ultimately execute. Section~\ref{sec:pl} describes
new directions in programming language abstractions and programmatic control over
networks that might ultimately enable these capabilities.

\paragraph{Performing automated, real-time inference:} The past ten years has demonstrated
significant promise in using
machine learning to both detect and predict network attacks; we must build on
the increasing amount of work in automated inference in network management
ultimately integrating it into a control loop that can enable more automated
decision-making. Section~\ref{sec:self-driving} describes two facets of these challenges:
(1)~using learning to improve network management and (2)~designing the network to
improve the quality of data that provides the input to learning algorithms. In a self-driving network, Quality of Data (QoD) is a prerequisite for quality of service (QoS) and, ultimately, the user's quality of experience (QoE).

\paragraph{Operating scalably in the data plane:} The networking community
has begun to lay the foundation for this aspect,
through fully programmable protocol-independent data planes (\eg, the Barefoot Tofino chipset~\cite{tofino} and Netronome NICs~\cite{netronome})
and the languages to program them (\eg, P4~\cite{p4}). Through these advances, data planes
are now beginning to support in-band measurements; coupled with distributed streaming
analytics platforms, there is huge potential for programmatic network control, not
only over forwarding (as SDN has enabled) but also over the collection of measurement data. Section~\ref{sec:data-plane} describes research challenges
and opportunities in these areas.

Operators have long wished for networks that are easier to manage;
developments in algorithms, machine learning, formal methods, programming
languages, and hardware design encourage us to think about the larger
goal of relieving the operator's burden as much
as possible, and possibly altogether. Indeed, the tools and technologies that could
help us realize
these goals are emerging, but even the pieces of the puzzle are not
complete: for example, the needs for automated control or inference place new
requirements on machine learning algorithms. A self-driving network thus
represents a grand challenge both for networking and broadly for computer
science. As we come to depend on the Internet for nearly everything
we do, it is a grand challenge we must undertake.

\section{Planning the Trip}
\label{sec:pl}

The first component of a self-driving network is {\em planning}, whereby a
network operator specifies high-level policies and a run-time system generates
corresponding measurement, inference, and control operations. Self-driving
networks should rely on a unified framework for specifying SLAs, network-wide
resource  optimization, and packet transformations and a runtime that
can generate the distributed programs that run on a heterogeneous collection
of network devices to integrate measurement, inference, and control.

\subsection{Specify sophisticated network policies} 

We envision a network whereby a network operator can specify (1)~the customer
expectations (\eg, statistical guarantees on latency and jitter); (2)~network-wide goals (\eg, minimizing congestion); and (3)~application
functions and services (\eg, network address translation, access control, intrusion
detection) that the network should satisfy.

\paragraph{Customer expectations (service-level agreements).} Network
operators should specify service-level agreements (SLAs) in terms
of guarantees on network metrics (\eg, latency, jitter, and DDoS response
time) or user quality-of-experience metrics, such as Mean Opinion
Score (MOS) for VoIP traffic or page load time for web browsing.  
Each SLA should correspond
to a particular subset of
traffic, specified by a predicate on packet header fields---or, better yet, on
higher-level names of Web sites (\eg, www.netflix.com) or applications (\eg,
video streaming)---and locations.  Interactive applications could
be assured that packet delay will be less than 10~msec at least 99.9\% of the time.
SLAs may correspond to contractual agreement with customers and can drive \emph{monitoring} (to detect when the network is at risk of
violating the guarantee), \emph{adaptation} (to alleviate the problem in the short term), and \emph{learning} (to ``learn" how to select configurations that satisfy SLAs without underutilizing the network). Today,  service providers specify SLAs informally.  Although some preliminary research presents languages for specifying SLAs~\cite{slang,sla-survey}, these works stop short of ``closing the loop'' on automatic monitoring, adapting, and learning.

\paragraph{Network goals (resource optimization).} In addition to satisfying
SLAs for customers, network operators aim to satisfy network-wide goals for running
their networks efficiently and reliably.  These goals can be naturally
expressed as optimization problems, with \emph{objectives} (such as minimizing
congestion) and \emph{constraints} (such as traffic conservation or limits on
path length).  Administrators should be able to specify these goals directly
as optimization problems.  For example, a common traffic engineering objective
is to minimize a sum over all links of some convex function $f()$ of link
utilization (\eg, $\sum_{\ell} f(u_{\ell}/c_{\ell})$) where link utilization
depends on the traffic matrix ($v_{ij}$, the volume of offered load from
ingress $i$ to egress $j$) and the routing ($r_{ij\ell}$), the fraction of
traffic from ingress $i$ to egress $j$ that traverses link $\ell$).  That is,
link utilization is the sum of all parts of the traffic matrix that follow
paths traversing the link (i.e., $u_{\ell}=\sum_{ij} r_{ij\ell}*v_{ij}$).  The
network operator should merely need to specify the objective function and
constraint---or, better yet, select these from a library of options---rather than
configure
traffic measurement and routing directly.  Rather than using a separate
traffic engineering tool, the network operator should be able to specify these
optimization goals in an integrated framework with other policy goals (\eg, SLAs and application services). Recent works~\cite{sol,merlin} take important steps in this direction, but stop short of integrating SLAs or automatically driving network measurement and inference decisions.

\paragraph{Services and functions (traffic transformations).} Network policies
go beyond quantitative measures of load, performance, and reliability, to
include operations performed on individual packets.  Network policies may
involve various packet transformations, including network address translation
and access control, as well as operations on packet payload (\eg,
transcoding and encryption).  In today's networks, these operations taking
place on specific {middleboxes}, which requires an operator to think at a ``box
level'', as opposed to specifying broader network goals. Network
operators should be able to specify traffic transformations
at a high level and have a runtime distribute the operations over network
elements, which may range from network switches to software virtual machines to
servers with hardware accelerators for specific operations.
Researchers have recently developed high-level languages for specifying
transformations of packets based on header fields and their
locations~\cite{hsa,veriflow,netkat}, including recent work on stateful
operations~\cite{snap-mina,stateful-netkat}. Some recent work also shows how
to synthesize a distributed configuration of network devices (\eg, OpenFlow or
P4 switches) to realize these policies while considering network-wide
optimization goals~\cite{merlin,snap-mina}. These developments are important building
blocks for the more ambitious goal of a self-driving network, which also entails
(among other challenges)
(1)~specifying a wider range of transformations
that operate on
packet payload or across packet boundaries (\eg, transcoding, compression, and
encrypt); and (2)~automatically ``compiling'' these specifications to a heterogeneous
collection of
network devices.

\subsection{Drive measurement, inference, \& control}

The run-time of a self-driving network should automatically generate
{both} the measurement queries and the control operations from a single high-
level specification, rather than requiring network operators to specify measurement
and control separately.  The run-time system should realize the combined
functionality directly in the data plane whenever possible. Below, we outline three example
scenarios that can benefit from tighter integration of measurement and control.

\paragraph{Minimizing network-wide congestion.} 
A policy could specify an optimization goal of minimizing network-wide congestion, as a sum over all links of a convex function of link utilization ($u_{\ell}$).  Link utilization is itself a function of the network routes and the traffic matrix.  Given such a specification, the runtime should automatically determine that it needs to measure the traffic matrix ($v_{ij}$ and configure the routing ($r_{ij\ell}$) by solving an optimization problem. In practice, the runtime must decide how often to collect the measurements (and to what degree of accuracy), how often to change routing (and how to minimize churn), and how to represent routing decisions (based on the capabilities of the network devices).

\paragraph{Shifting traffic from congested peering point.}
When traffic on a particular peering link exceeds a threshold, an operator might
want
excess traffic to spill over to a secondary interconnection link.  Based on this
policy, the runtime
system should
monitor traffic load on the first link and decide whether and how
to balance traffic load.  Rather than relying on a static threshold,
the decision might also rely on
a higher-level QoE metric (such as MOS, or even direct signaling from an application
about video bitrates or rebuffering) that triggers monitoring of QoE for the associated
traffic.

\paragraph{Detecting and blocking unwanted traffic.}
An operator might outline a policy to detect and mitigate denial-of-service
(DoS) attacks; the policy might specify that the network should rate
limit traffic sent to a destination receiving a particular type of DNS
response message from many distinct senders.  Based on this policy, the
runtime should generate the necessary monitoring queries
and, based on the monitoring results, rate-limit the suspicious traffic.
Rather than detecting DoS attacks using specific thresholds, the policy could
specify a detection technique (\eg, sequential hypothesis testing
for port-scan detection) for identifying  attacks.

\section{Navigating in a Dynamic Environment}
\label{sec:self-driving}

The network's complexity and the dynamic nature of its underlying
processes make machine learning algorithms a natural tool for detecting,
diagnosing, and mitigating disruptions. Previous work has applied
techniques from both machine learning and user interaction to improve
specific aspects of network
security~\cite{antonakakis2010,hao2009,gu2008} and
performance~\cite{tariq2008,li2010}. To date, however, these techniques
have been primarily ``bolted on'' to existing designs, rather than
incorporated directly into the network's control fabric. For example,
many applications of machine learning to network security have involved
development and (often offline) testing of algorithms with bulk traffic
traces; the next natural step is to integrate these types of inference
and control algorithms into the network's decision and control fabric.
Even applying existing learning algorithms has often proven difficult,
partially because existing network protocols and technologies do not
make it easy to obtain labels for data samples. Conversely, today's
machine learning algorithms are often not tailored for network data,
which is high-volume, distributed, and rapidly evolving; existing
algorithms also make it difficult to iteratively refine the features
used in a supervised learning algorithm (as might be required for
high-volume network traffic traces) or to perform complex timeseries
analysis.

In this section, we describe how techniques and insights
from both machine learning and user interaction can help facilitate self-driving
networks by:
(1)~incorporating machine learning-based inference into the network so that, in
many cases, the network can learn to run itself, removing many of the decisions
from network operators (Section~\ref{sec:ntl}); (2)~incorporating input from applications
and human users to better improve
the inputs to learning algorithms (Section~\ref{sec:hcn}).

\subsection{Improving operations with learning}\label{sec:ntl}

Networks should provide high availability, good application performance, and
security in the face of disruptions using automated and semi-automated
introspection. In contrast to past approaches, which patch the existing
network with point solutions (\eg, middleboxes such as firewalls and spam filtering
appliances), we propose to make the functions provided by these boxes inherent
to the network itself. We will
discuss two areas of network management that are amenable to self-driving operation:
(1)~satisfying performance requirements and
service-level agreements;  and (2)~automated detection and mitigation of
unwanted traffic (\eg, spam, DoS attacks).

\paragraph{Performance: Applications and service-level agreements.} 
Providing good network performance involves both reacting to changing network
conditions on short timescales. Providing good network performance for some
application requires understanding the relationships between application-level
metrics (\eg, video bitrate, rebuffering events) and what can be measured from
traffic as it traverses the middle of the network. In other cases, an
operator's task may involve a contractual service-level agreement
(SLA)---including determining when network conditions might cause an SLA to be
violated.  When networks were simpler, it was possible to model the behavior
of (say) a TCP connection using closed-form analysis, as well as to predict
how certain network changes (\eg, the change of routing protocol weights)
might affect the performance of an application. In today's networks, however,
this type of closed-form analysis is no longer tractable, largely to the
complexity of deployed networks and the many interacting network components
that collectively contribute to the performance of the network and 
applications.

With the ability to collect, store, and
analyze additional data, networks can produce models that establish more
complex relationships between lower-level metrics such as utilization and
higher-level metrics such as streaming application performance. For example,
previous work has established that it is possible to model how specific
provisioning decisions ultimately affect web search response
time~\cite{tariq2008}. Past work has demonstrated that it is possible to learn
relationships between lower-level network features (\eg, round-trip latency)
and application performance metrics  (\eg, search response time). Developments
in the speed and sophistication of these algorithms, coupled with advances in
data-plane programmability, suggest that we should think about extending these
techniques to problems concerning monitoring and control over real-time
performance, including application-level performance guarantees and SLA
monitoring.

\paragraph{Security: Unwanted traffic.} Recent years have seen
significant advances in applications of machine learning to statistical
anomaly detection.  Research has developed learning algorithms to detect
(and even predict) attacks based on analysis of network traffic (from
packet traces to IPFIX records)~\cite{lakhina2004pca}, DNS
queries~\cite{antonakakis2010} and domain registrations~\cite{hao2016},
and even BGP routing messages~\cite{konte2015}.  Yet, most of these
anomaly detection algorithms have only been demonstrated on {\em
  offline} traffic traces; such demonstrations are useful for
identifying features for anomaly detection algorithms that run on
stand-alone network appliances; the prospect of a self-driving network
raises many more challenges and opportunities. One challenge involves
tailoring these algorithms to operate in real time, coupled with
real-time action. For example, simple regression models based on
lightweight features could be executed in programmable switches that
support customizable feature extraction and computation (\eg, those
based on the Barefoot Tofino chipset~\cite{tofino}); we discuss this challenge further
in Section~\ref{sec:ml-data}. An additional challenge involves
developing a new class of machine learning algorithms whereby an
algorithm could perform an initial rough classification based on
lightweight features (\eg, those based on metadata or coarse statistics)
and trigger collection of more heavyweight features (\eg, those from
packets) when classification is uncertain; we explore this possibility
in more detail in the next section.

\subsection{Improving learning with better data}\label{sec:hcn}

Networks should also be tailored to improve the quality of input data provided
to real-time inference and prediction algorithms. For example, machine
learning algorithms for network security such as intrusion detection often
train on labeled data. Yet, for the domain of network security, obtaining
labeled data is difficult: attacks are rare, threats are dynamic, and new
classes of threats and attacks are continually emerging. Similarly,
identifying quality of experience degradations often requires input from
applications, users, or both. In this section, we discuss how future networks might
be {\em co-designed} with learning algorithms to improve algorithm accuracy, and
to improve the quality and quantity of data that provides input to these algorithms.

\subsubsection{Improving model accuracy}

\paragraph{Input from high-level policy and topology dependencies.}
Conventional machine learning methods operate on offline network traces, with little to no information about a network's structural
dependencies and, as a result, must infer much of what is already known before it can make any useful inferences.  New machine learning
techniques might better diagnose network problems by incorporating input from the network topology (\eg, shared risk link groups) and the high-level policy. Consider
the case of detecting network faults that affect 
availability. Unfortunately, although networks offer a wealth of data,
they lack a
single framework that synthesizes heterogeneous data to form hypotheses about
underlying causes. For example, the failure of a single link can cause link alarms, routing changes, and traffic shifts.  Rather than forcing a machine-learning algorithm to infer these dependencies from observations of failure events, the self-driving network can draw on information about the network topology.

\paragraph{Collecting additional data to improve model accuracy.} The accuracy
of an inference model may also depend on the type and quantity of data that is
available. In many cases, inference algorithms improve with additional data
samples, or data of a different type or granularity. A network that learns
could use a coarse detection algorithm based on network data that is
relatively lightweight or easy to collect (\eg, sampled IPFIX logs, SNMP) to
develop a classifier that might have a false positive rate that is higher than
acceptable. The output of this classifier might trigger additional
measurements---either active measurements (\eg, probes) to and from different
parts of the network or, in some cases, more expensive packet captures that
could provide more precise information about the traffic (\eg, DNS query logs,
timing information). The emergence of technologies such as in-band network telemetry~\cite{INT}
make it possible not only to write additional fine-grained information into packets,
but also to {\em generate} probe traffic on demand, making it possible to trigger
fine-grained active and passive measurements either end-to-end or from within the
network, should an algorithm need that information.

\subsubsection{Improving data quality}

\paragraph{Increasing the amount of labeled data.} One of the challenges in
applying machine learning to network performance and security problems is the
paucity of labeled data with which to train these algorithms. The lack of
labeled data is fundamental to today's networks, for several reasons: Many
interesting events are (1)~rare (\ie, they do not happen frequently enough to
generate a reasonable training set); (2)~emerging (i.e., they reflect a new
class of threat or attack that was previously unseen); or (3)~dynamic. In the
case of network faults or failures, examples are rare: When a network fault
occurs, it is often due to a ``one off'' misconfiguration of a network device
that is interacting with other devices on the network in unexpected (and
previously unobserved) ways. Other network faults may occur when physical
hardware fails or when a particular traffic pattern tickles an
implementation bug or configuration error. Unfortunately, because each failure
is essentially unique, training based on past examples of failures may not
produce a classifier that can detect and diagnose future failures. A network
that learns could incorporate information directly from operators, from
network configuration, or perhaps even from users or applications
to increase the amount of labeled data that detection
and inference algorithms could use to train.

\paragraph{Input from users.}  Feedback from end users can help drive
additional passive and active measurements in the network. Network operators
typically have visibility into metrics in the network itself, but these
metrics are sometimes difficult to map to user experience. We envision that
these vantage points might be better coupled through explicit feedback from
users that could subsequently trigger additional passive or active
measurements. One possibility, for example, is that applications such as a Web
browser have a button whereby users could explicitly indicate poor application
performance (an ``I'm frustrated'' button). This feedback could result in
annotations on packets in application traffic that could trigger additional
passive or active measurements from switches.

Application developers occasionally poll
end users about the performance of individual applications (\eg, ``How was
your experience on the last video call?'') through a technique known as
experience sampling. One challenge associated with experience sampling
concerns when to poll users about their experiences: infrequent sampling can
result in inadequate data about application performance; on the other hand,
sampling that is too frequent risks irritating the user or causing the user to
submit dismissive responses. One possible line of research is to use network
measurements to drive and automate experience sampling. For example, a
programmable switch in the network or an instrumented OS kernel might indicate
a degradation in conditions, such as higher packet loss or latency, or a
reduction in throughput; similarly, a server might be able to witness elevated
packet loss or latency in a TCP stream. These conditions could serve as
automated triggers for polling a user about application experience; with the
appropriate integration, a network device or server could generate a packet
that could be automatically parsed by the user's operating system or browser
to trigger the sample.

\paragraph{Input from applications and operating systems.} End-user
applications often have precise information about the performance they are
experiencing (\eg, whether a rebuffer event occurred, the fact that the video
bitrate changed) but often have no way of communicating this information to
the network. Similarly, the operating system may have additional information
about user engagement, such as whether an application is running in the
foreground and perhaps even whether a user is engaging with the application
(or device!) at all. Communicating information both about application
performance and user engagement to the network could facilitate more efficient
use of network resources.  An operating system could include signaling
information about application state into network traffic flows, which the
network could subsequently use to assign the traffic to a higher or lower
priority queue. Such a capability could be useful, for example, if the network
could determine that it could safely de-prioritize a high-throughput video
stream that the user was no longer watching, even though the video continued
to stream. Additional information from applications and operating systems, such
as TCP statistics, could also be used to label traffic streams that could later
be used as attributes in queries.

\section{The Need for Speed}
\label{sec:data-plane}

The capabilities in previous sections rely on real-time monitoring and prediction, streaming
analytics on high-volumes of network traffic, and line-rate processing
functions ranging from simple functions such as aggregation to more complex
functions like inference and prediction. Many research challenges lie ahead,
in both designing and applying these building blocks for self-driving networks---particularly 
in making these functions scalable, distributed, and real-time.

\subsection{Traffic analytics in the data plane}

Flexible packet parsing, match-action pipelines, and the ability to
maintain state both in the switch and on packet headers can enable
networks to support high-level measurement abstractions.

\paragraph{Compact data structures.}  Programmable switches can perform
arithmetic
operations and maintain state in tables, allowing switches to support compact
data structures that maintain statistics about packet streams.
These data structures can support higher-level
abstractions such as maintaining sets (\eg, Bloom filters), counts (\eg,
counting Bloom filter or count-min sketch), or counts of unique items (\eg,
count distinct sketch). Recent studies have shown how to support these kinds
of data structures on emerging switches~\cite{OpenSketch, UnivMon,
HashPipe}; more work lies ahead in optimizing for limited
state, computational resources, and control bandwidth.

\paragraph{Piggybacking state on packets.}  Many networking tasks require
operations across multiple hops. The ability to tag packets with
state and update that state at subsequent hops
enables the data plane to support a range of powerful abstractions.  For example,
a packet header could carry the version of network policy applied to that packet
(\eg, to support consistent policy updates~\cite{consistent-updates}, sets or sequences (\eg, of next-hops, network paths, or
middleboxes for flexible traffic steering\cite{r-encoding}); states of a
deterministic finite automaton to evaluate a regular expression on the
properties of a packet and its path through the network, to measure or control
traffic based on these properties~\cite{path-queries}; or the aggregation of
traffic statistics across a path, to collect path-level metrics such as
maximum link utilization or total queuing delay~\cite{INT}. 

\paragraph{Simplifying joins with other datasets.}  Analysis often requires
joining traffic statistics with other data sets. For example, joins can
associate a packet's destination address with its autonomous system (by
joining with routing table data), website or application (by joining with DNS
query logs), or the end user (by joining with authentication server data).
This information can facilitate aggregation of measurement data, the routing
and scheduling of traffic, and access control, based on higher-level policies.
In today's networks, these joins are cumbersome, often relying on coarse-grained timestamps from different locations. The data plane can simplify the
join process in two ways.  First, the data plane can perform the join itself
by analyzing and combining  datasets simultaneously or by maintaining an
efficient representation of the second dataset (\eg, a table of IP addresses
associated with authenticated users in a particular class).  Second, a switch
can tag packets representing, for example, a location in the network and
associated timestamp, with information that can simplify a subsequent join.

\subsection{Prediction models in the data plane}\label{sec:ml-data}

As discussed in Section~\ref{sec:self-driving}, machine learning has been applied
to a
wide variety of network monitoring tasks, ranging from performance monitoring
to security. To date, however, many of these models have been demonstrated and
deployed in a purely offline fashion: Traffic is collected from the network in
the form of packet captures, IPFIX records, or DNS query logs and is used to
train a detection model, which is also evaluated offline. Yet, many of
these models incorporate simple features---often ones that can be computed
or inferred from a single packet. Programmable switches could extract these
features from the packets in the data plane and even compute regression
functions based on these learned models, essentially computing the prediction
function in-line and making real-time decisions about the nature of traffic in
the network, without ever requiring off-path analysis.

Consider a machine-learning based spam filter based on network-level features
such as the autonomous system of the sending IP address, and the number of
adjacent IP addresses that have also sent emails~\cite{hao2009}. Programmable
switches could compute these features inline {\em and} compute the weighted
linear combination of individual feature characteristics to compute the
overall likelihood that a message is spam. Another example involves botnet
detection based on DNS lookups: these classifiers detect abnormal features
such as lookups that are (among other features) lexicographically close
together, occur in large bursts over short time intervals, and are hosted on
authoritative DNS servers with known bad reputations.  A programmable switch
could parse the DNS queries to extract these features and detect DNS lookups
associated with malicious activity in the switches themselves, without ever
requiring offline analysis.

\balance\section{Conclusion}\label{sec:conclusion}

The increasing performance, reliability, availability, and security demands of
modern networked applications are making network management more important
than ever. At the same time, networks themselves have become far too complex
to manage using state-of-the-art approaches, which rely on closed form models
of network behavior and performance at the level of individual protocols and
devices. As a community, we must consider a fundamentally new approach to
network management that (1)~relies instead on data-driven models that can
predict end-to-end network performance from lower-level metrics; (2)~couples measurement
with real-time control, eliminating the operator from the management control loop
whenever possible. The past decade has laid the groundwork for designing networks
that drive themselves, with technologies ranging from statistical anomaly detection
and learning-based troubleshooting tools to programmable networks and compact data
structures for line-rate algorithmics. We should aspire to use these building blocks
to build the self-driving networks that our applications now demand.
\label{lastpage}

\end{sloppypar}


\pagebreak

\small
\setlength{\parskip}{-1pt}
\setlength{\itemsep}{-1pt}
\balance\bibliography{paper}
\bibliographystyle{abbrv}
}{
}

\end{document}